\documentclass[prl,aps,amssymb,twocolumn,noshowpacs,hyperref]{revtex4-1}
\usepackage{color}
\usepackage{graphicx}
\usepackage{hyperref}
\usepackage{amsmath}
\usepackage{latexsym}
\usepackage{amssymb}
\usepackage{mathrsfs}
\usepackage{mathtools}
\usepackage{layout}
\usepackage{verbatim}
\usepackage{multirow}
\usepackage{bm}
\usepackage{amsfonts,epsfig}
\usepackage{lineno}
\usepackage{textcomp}

\begin{document}
\title{Fast nuclear-spin entangling gates compatible with large-scale atomic arrays}

\date{\today}
\author{Xiao-Feng Shi}
\affiliation{School of Physics, Xidian University, Xi'an 710071, China}
\author{Yan Lu}
\affiliation{School of Physics, Xidian University, Xi'an 710071, China}

\begin{abstract}
Nuclear-spin entangling gates with divalent atoms can be executed by one global laser pulse when $\Delta_{\text{Z}}<\Omega$, where $\Delta_{\text{Z}}$ is the Zeeman-splitting-dominated frequency difference for the clock-Rydberg transitions of the two nuclear-spin qubit states and $\Omega$ is the maximal Rabi frequency. Concerning the sensitivity of Rydberg-state energy to magnetic fluctuation, the gate is compatible with large-scale atomic arrays for weaker magnetic field is suitable for ensuring uniform field in a large qubit array. The gate can have a high fidelity because the relaxation and dephasing of Rydberg states, which limit the fidelity and grow with $1/\Omega$, can be mitigated with easily attainable large $\Omega$.
\end{abstract}
\maketitle

\section{introduction}\label{sec01}
Quantum computing requires both scalability and accuracy~\cite{Nielsen2000,Williams2011,Ladd2010}, for which several physical platforms have demonstrated high-fidelity control on the single quantum level over several tens of qubits~\cite{IonQ,Evered2023,Acharya2023}, yet it remains prohibitive to realize both a large-scale qubit array and universal quantum logic gates executable with qubits in any location of the array.

It was recently found that individual neutral atoms can form a quantum memory with hundreds of identical qubits~\cite{Schymik2020,Ebadi2021,Pause2023}, but scaling nuetral-atom qubit array without hampering its controllability is not straightforward though qubit arrays over 1000 atoms~\cite{Pause2023}, high-fidelity single-qubit~\cite{Olmschenk2010,Xia2015,Wang2016,PhysRevLett.131.030602} and two-qubit gates~\cite{Evered2023,Scholl2023,Ma2023}, and small-scale quantum processors~\cite{Graham2022,Bluvstein2022} were already demonstrated with neutral atoms. The headache partly arises from two aspects. {\it First}, for an alkali-metal atomic qubit array as in ~\cite{Wilk2010,Isenhower2010,Zhang2010,Maller2015,Jau2015,Zeng2017,Levine2018,Picken2018,Levine2019,Graham2019,Jo2019,Fu2022,McDonnell2022,Bluvstein2022,Graham2022,Evered2023}, the spatial change~\cite{footnote2}
of magnetic field $B\mathbf{z}$ in an anticipated large-scale array with practically useful millions of qubits~\cite{Saffman2019nsr} can result in spatial change of frequency separation of the two qubit states, which effectively makes the qubits no longer identical. {\it Second}, dominant error in a Rydberg-mediated entangling gate~\cite{Wilk2010,Isenhower2010,Zhang2010,Maller2015,Jau2015,Zeng2017,Levine2018,Picken2018,Levine2019,Graham2019,Jo2019,Fu2022,McDonnell2022,Bluvstein2022,Graham2022,Evered2023} is from the dephasing of the ground-Rydberg transition, decay of Rydberg state, atomic position fluctuation, and scattering at the intermediate state~(with alkali-metal atoms)~\cite{Graham2019,Evered2023}, all grow with the duration of Rydberg excitation pulse. Effective suppression of these errors can be via large Rydberg Rabi frequency $\Omega$. For alkali-metal atoms, unfortunately, an intermediate state was often employed~\cite{Wilk2010,Isenhower2010,Zhang2010,Maller2015,Zeng2017,Levine2018,Picken2018,Levine2019,Graham2019,Jo2019,Fu2022,McDonnell2022,Bluvstein2022,Graham2022,Evered2023} at which the scattering strongly limits $\Omega$~\cite{Evered2023}.

One solution to the above issues is encoding qubits by nuclear spins in the clock state of alkaline-earth-like atoms. {\it First}, even with spin-orbit and hyperfine interaction induced state mixing~\cite{Boyd2007}, the clock state has a g factor of the nuclear-spin character~\cite{footnote1} which renders a sensitivity to magnetic field that is three orders of magnitude weaker than that of a hyperfine qubit in alkali-metal atoms, therefore nuclear-spin qubits are more compatible with large-scale qubit arrays. {\it Second}, Rydberg excitation of the clock state of an alkaline-earth-like atom via no intermediate states with $\Omega>2\pi\times10$ is much easier~\cite{Madjarov2020} compared to that of an alkali-metal atom~\cite{Hankin2014}, thereby can potentially enhance the gate fidelity since the intermediate state scattering is absent and Rydberg-state decay can be suppressed with much shorter gate durations. However, using entangling protocols designed for alkali-metal atoms~\cite{Shi2019prap,Levine2019,Jandura2022} with nuclear-spin qubits requires $\Delta_{\text{Z}}\gg\Omega$~(e.g., the experiments in~\cite{Ma2022,Ma2023} had $\Delta_{\text{Z}}/\Omega$ equal to 25 and 5.8, respectively), where $\Delta_{\text{Z}}$ is the Zeeman-splitting-dominated detuning for the Rydberg excitation of the two nuclear-spin qubit states. Gates in the condition $\Delta_{\text{Z}}\gg\Omega$ will be slow under a Gauss-scale B-field, or magnetic noise can be significant if strong B-fields are used for fast gates. It is a demanding task to find a fast gate with $\Delta_{\text{Z}}/\Omega<1$ when meanwhile the gate is experimentally friendly, namely, needs only one global laser pulse~\cite{Evered2023,Ma2023}.   

Here, with qubits encoded in the clock states of alkaline-earth-like atoms so that large $\Omega$ is realizable~\cite{Madjarov2020}, we study entangling gates realized under the condition $\Delta_{\text{Z}}/\Omega<1$ via one ultra-violet laser pulse. The short gate duration and absence of intermediate-state scattering can help to yield high-fidelity gates for $\Omega$ over $2\pi\times10$~MHz is readily attainable~\cite{Madjarov2020}. Importantly, a Gauss-scale B-field can be used without compromising the gate speed, thus the gate is compatible with a large atomic array because in order to have approximately the same B-field throughout the whole atomic array, the array must be placed, e.g., near the center of an exceedingly large solenoid where the large size decreases the B-field.

\section{A CZ-like quantum gate}
\begin{figure}
\includegraphics[width=3.4in]
{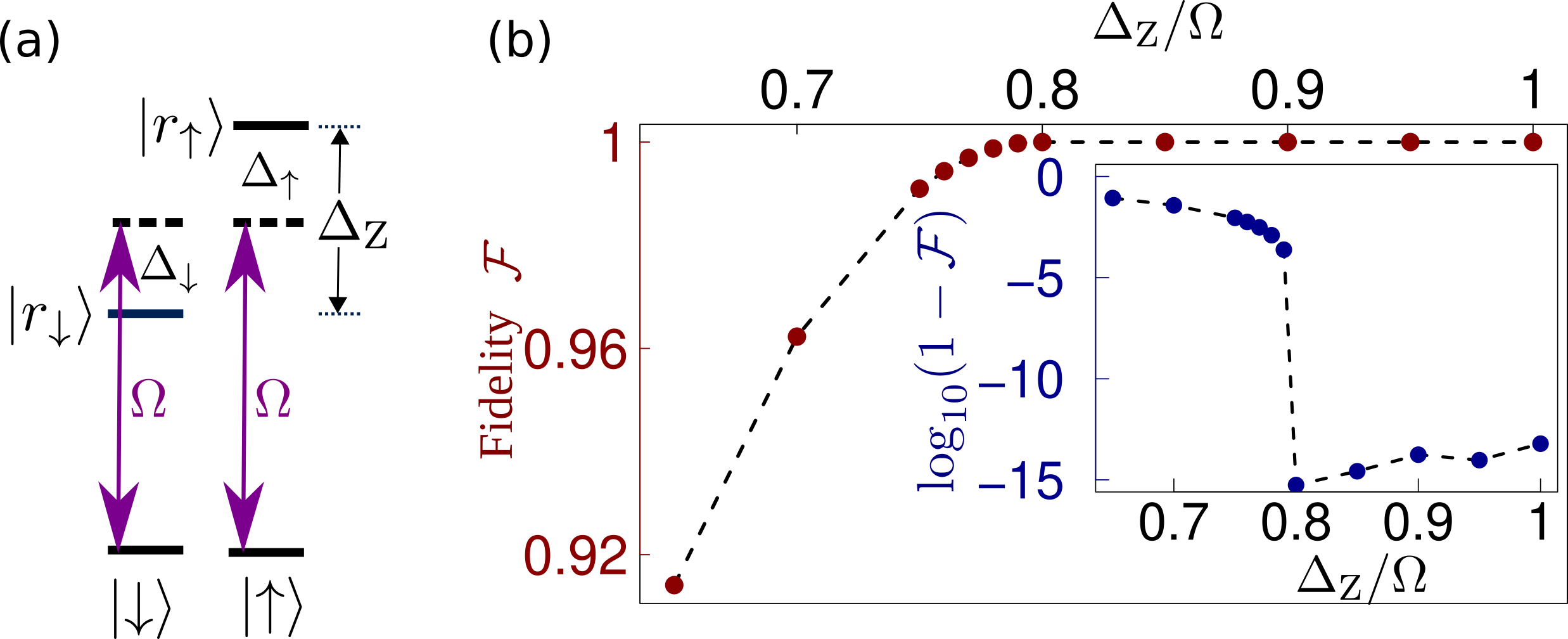}
\caption{(a) A $\pi$-polarized ultra-violet laser sent to, as an example, a $^{171}$Yb atom, exciting two nuclear-spin qubit states $\lvert \uparrow\rangle\equiv\lvert (6s^2)^3P_0 \lvert I=1/2, m_I=1/2\rangle$ and $\lvert\downarrow\rangle\equiv \lvert (6s^2)^3P_0 \lvert I=1/2, m_I=-1/2\rangle$ to two respective Rydberg states $\lvert r_\uparrow\rangle,\lvert r_\downarrow\rangle\equiv (6sns)~^3S_1 \lvert F=3/2, m_F=\pm 1/2 \rangle$ with a Rabi frequency $\Omega$.  (b) By assuming perfect blockade and with a pulse duration $3\pi/\Omega$, a CZ-like gate in Eq.~(\ref{Gate01}) is realized by smoothly changing the phase of the laser. Shown is the gate fidelity $\mathcal{F}$ and the common logarithm of the infidelity in the inset.  \label{figure1} }
\end{figure}

The CZ-like quantum logic gate in this paper is realized by phase accumulation of two-qubit nuclear-spin states in detuned Rydberg excitation~\cite{Shi2017,Shi2019prap,Levine2019} in the blockade regime~\cite{PhysRevLett.85.2208,Lukin2001}. With qubits encoded in two nuclear spin states of the clock state of an alkaline-earth-like atom such as $^{171}$Yb~\cite{Ma2023}, $\lvert\uparrow(\downarrow)\rangle\equiv  (6s6p)~^3P_0\lvert m_I=\pm1/2 \rangle$, the gate maps the four computational basis states as
\begin{eqnarray}
 &&\lvert  \uparrow  \uparrow\rangle \rightarrowtail{ } e^{i\alpha}\lvert  \uparrow  \uparrow\rangle,\nonumber\\
&& \lvert  \uparrow  \downarrow\rangle \rightarrowtail e^{i(\alpha+\beta)/2} \lvert  \uparrow  \downarrow\rangle ,\nonumber\\
&& \lvert  \downarrow  \uparrow\rangle \rightarrowtail e^{i(\alpha+\beta)/2} \lvert  \downarrow  \uparrow\rangle,\nonumber\\
&& \lvert  \downarrow  \downarrow\rangle \rightarrowtail -e^{i\beta } \lvert  \downarrow  \downarrow\rangle,\label{Gate01}
\end{eqnarray}
which transforms to the canonical controlled-Z~(CZ) gate by single-qubit gates $\{\lvert   \uparrow\rangle, \lvert   \downarrow\rangle\}\rightarrowtail{ }  \{ e^{-i\alpha/2} \lvert  \uparrow\rangle, e^{-i\beta/2} \lvert   \downarrow\rangle\}$. The two qubit states in each atom are nearly degenerate in a Gauss-scale magnetic field $B\mathbf{z}$, but during the excitation to two hyperfine-Zeeman substates $\lvert r_{\uparrow,\downarrow}\rangle$ of a $(6s6n)~^3S_1$ Rydberg state with one laser field~\cite{Shi2021,Shi2021pra,Chen2022,Shi2024}, the two nuclear-spin states have a detuning difference $\Delta_{\text{Z}}\equiv|\Delta_{\uparrow}-\Delta_{\downarrow}|$ which is approximately the Zeeman splitting $\Delta_{\text{Z}}\approx2\pi\times1.9B$~MHz$/$G~\cite{Chen2022}between $\lvert r_{\uparrow}\rangle$ and $\lvert r_{\downarrow}\rangle$~\cite{Shi2021,Shi2021pra,Chen2022,Ma2022,Ma2023,Shi2024}, where $\Delta_{\uparrow,\downarrow}$ is the detuning of the laser field with respect to $\lvert r_{\uparrow,\downarrow}\rangle$ shown in Fig.~\ref{figure1}(a). During the clock-Rydberg excitation when $\Delta_{\uparrow,\downarrow}$ is comparable to the Rabi frequency $\Omega$, a significant phase can arise when both $\Delta_{\uparrow,\downarrow}$ and $\Omega$ are fixed~\cite{Shi2017,Shi2024}. But this makes a simultaneous restoration of the four computational basis states back to themselves impossible unless multiple laser pulses are used ~\cite{Shi2024}, so we consider a smooth change of laser phase~\cite{Jandura2022,Evered2023,Ma2023}. A smooth change of laser frequency can also be used, but it can lead to more population leakage via undesired transitions due to the polarization impurity of the laser field. Therefore we focus on a smooth change of laser phase and $\Delta_{\text{Z}}/2=-\Delta_{\uparrow}=\Delta_{\downarrow}$ so that undesired transitions due to polarization impurity of laser fields are more detuned as discussed later. In this work we focus on linearly polarized laser field as often used in experiments~\cite{Ma2023}.

\begin{table}[ht]
  \centering
  \begin{tabular}{|c|c|c|c|c|c|c|c|c|c|}
    \hline
 $\frac{\Delta_{\text{Z}}}{\Omega}$ & 0.6 &0.65 &0.7&0.75&0.8& 0.85  & 0.9 &0.95&1\\
   \hline
 $N$  & 1.843 &  1.733&1.646& 1.568 &1.497 & 1.434    &1.376   & 1.325& 1.291\\
   \hline
  \end{tabular}
  \caption{$N$ on the second row indicates the minimal gate duration $2\pi N/\Omega $ required to realize a CZ-like gate in Eq.~(\ref{Gate01}) as a function of $\Delta_{\text{Z}}/\Omega$ when the infidelity~\cite{Pedersen2007} can be smaller than $10^{-7}$ when assuming infinite Rydberg blockade interaction in the numerical optimization~\cite{Khaneja2005,DeFouquieres2011} and absence of Rydberg-state decay. Here, $\Delta_{\text{Z}}\equiv|\Delta_{\uparrow}-\Delta_{\downarrow}|$ and change of ratio between $\Delta_\uparrow$ and $\Delta_\downarrow$ does not alter the results here. Too small $\Delta_{\text{Z}}$ doesn't work. For example, $\Delta_{\text{Z}}/\Omega=0.55$ leads to $\mathcal{F}\lesssim0.991$ when $N\leq1.9$, where we restrict the values of $N$ because the condition $N>1.9$ brings no advantage on the gate duration in units of $1/\Delta_{\text{Z}}$ compared to the cases here.    \label{table-0}  }
\end{table}

To study quantum gates compatible with large arrays, small B-field and gate time are favorable, and meanwhile the gate fidelity must  be large. In this work, we assume that the dipole-dipole interaction $V$ between the two qubits when both are in Rydberg states is infinite so that we can theoretically investigate how fast a gate can be under a certain B-field, or how small the B-field can be when we fix the gate duration. This is because $V/\Omega$ can be huge in the blockade regime~\cite{Saffman2010,Saffman2016,Weiss2017,Saffman2019nsr,Adams2020,Browaeys2020,Morgado2021,Wu2021,Kaufman2021,Shi2021qst}, and also because when we use a finite $V$ in the numerical simulation, dynamical phases can arise which depend on the value of $V$~\cite{Shi2024}, and optimal control~\cite{Saffman2020,PhysRevA.105.042404,Jandura2022,PhysRevApplied.18.044042,Jandura2023,PhysRevResearch.5.033052,Sola2023a,Sola2023b,Li2023,Sun2023,PhysRevLett.131.110601,HHJen2023,PRXQuantum.4.040333} can locate a $V$-dependent pulse so as to reach unit fidelity but it can be a biased prediction on how fast the gate can be.
\begin{figure}
\includegraphics[width=3.4in]
{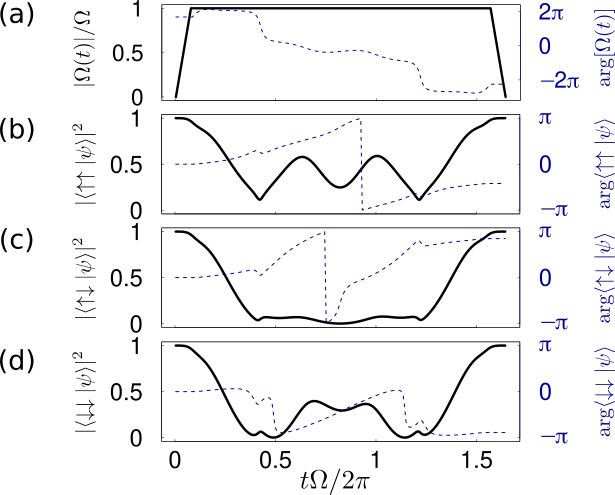}
\caption{State dynamics for the CZ-like gate with $\frac{\Delta_{\text{Z}}}{\Omega}=0.8,~\Delta_\downarrow=-\Delta_\uparrow=\Delta_{\text{Z}}/2$. (a) shows the amplitude and phase of $\Omega(t)$. (b,c,d) show the population~(solid curve) and phase~(dashed curve) of the ground-state component of the wavefunction when the input states are $\lvert  \uparrow  \uparrow\rangle,\lvert  \uparrow  \downarrow\rangle$, and $\lvert  \downarrow  \downarrow\rangle$, respectively; the state dynamics for $\lvert  \downarrow  \uparrow\rangle$ is similar to that of $\lvert    \uparrow\downarrow\rangle$. Rydberg-state decay and blockade leakage are ignored here. The Rydberg superposition time is about $2\pi/\Omega$ averaged over the four input states. Here the gate duration is 10$\%$ more than $2\pi N/\Omega $ where $N=1.497$ for we have added a rise and fall edge of $\Omega(t)$, each of duration $\pi N/(10\Omega)$. Concerning the simultaneous modulation of laser intensity and phase may be more challenging compared to the modulation of either intensity or phase, here the rise and fall edge of the laser field have constant phases.    \label{figure2} }
\end{figure}

\section{Limit of gate speed}
To find a gate with a fast speed so as to have a high fidelity, and meanwhile being compatible with a large array, we note that the value of $\Omega$ for the clock-Rydberg transition can be quite large~\cite{Madjarov2020}, but the CZ-like gate shall be executed with a $\Omega$ that has a certain ratio with $\Delta_{\text{Z}}$ as indicated in Table~\ref{table-0}, therefore $\Delta_{\text{Z}}$ is the key parameter limiting the gate speed in the regime $\Delta_{\text{Z}}/\Omega<1$. In order to have a high-fidelity gate with any atom pair in a large-scale atomic array, the B-field, or, $\Delta_{\text{Z}}$, should be small. To understand this, we can consider the qubit array with 1305 atoms in an about $30\mu$m$\times30\mu$m area reported in Ref.~\cite{Pause2023}. For a practically useful quantum computer with, e.g., a million atoms~\cite{Saffman2019nsr}, we would anticipate a scaling of the array in~\cite{Pause2023} to a larger one in a $900\mu$m$\times900\mu$m
area. If the coil to ensure a smooth magnetic field throughout the qubit array in Ref.~\cite{Pause2023} has a radius $30$~cm, then we would need another coil with a radius $9$~m for the million-qubit atomic array to ensure a similarly smooth magnetic field. Though sounds crazy, it is not a forbidden task and worthy concerning the benefits a practical quantum computer can bring to us~\cite{Shor1997,Grover1997,Grover1998}. But with such a large coil, it is in general not easy to generate a strong and stable B-field at the qubit array, so we can assume $B$ smaller than, say, $10$~G~\cite{Isenhower2010,Zhang2010,Wilk2010,Maller2015,Levine2018,Jau2015,Zeng2017,Picken2018,Graham2019,McDonnell2022,Levine2019,Bluvstein2022,Evered2023,Ma2022,Ma2023} at which $\Delta_{\text{Z}}/2\pi=19$~MHz in the case of $^{171}$Yb as an example. This assumption is made also because weaker B-fields have smaller fluctuation~\cite{Schine2022}. When we fix the gate duration with a given $\Omega$, there is a minimal magnetic field below which the gate in Eq.~(\ref{Gate01}) can't be realized, as shown in Fig.~\ref{figure1}(b). For example, with a gate duration $3\pi/\Omega$, pulses can be found to yield a unit fidelity when $\Delta_{\text{Z}}/\Omega\gtrsim0.8$. On the other hand, it is desirable to implement a gate within a short time $2\pi N/\Omega $~\cite{Jandura2022,Evered2023,Ma2023}. We find that the smallest $N$ is about 1.3 when $\Delta_{\text{Z}}/\Omega$ is 1, and when $\Delta_{\text{Z}}/\Omega$ decreases, $N$ grows. Moreover, a too small $\Delta_{\text{Z}}$ is not suitable for fast gate. With a decrease of 0.05 from $\Delta_{\text{Z}}/\Omega=1$, Table~\ref{table-0} shows that the smallest $\Delta_{\text{Z}}/\Omega$ is $0.6$ for realizing a gate in Eq.~(\ref{Gate01}), with a gate duration about $2.21\pi/\Delta_{\text{Z}}$, which is about 58~ns when $B=10$~G~(the gate with circular field can have a smaller duration, $1.8\pi/\Delta_{\text{Z}}$). In experiments, there can be a rise and fall edge of the laser field. With the rise and fall edge included, one can still have a pulse for the CZ-like gate, where one example is shown in Fig.~\ref{figure2}. The phase profile in Fig.~\ref{figure2}(a) is slightly different from that without rise and fall edge. The example of Fig.~\ref{figure2} has a Rydberg superposition time  $2\pi/\Omega$ leading to a Rydberg-state-decay induced gate error $1.6\pi/(\tau\Delta_{\text{Z}})$, which is about $4.2\times10^{-4}$ with $B=10$~G if $\tau=100~\mu$s~\cite{Shi2024}, i.e., the decay-induced error is greatly suppressed. Besides the suppression of the Rydberg-state decay, such a fast gate can also suppress the error from the motion-induced dephasing; in a recent experiment~\cite{Evered2023} a high-fidelity entangling gate was realized thanks to a short gate duration less than $260$~ns.

\begin{figure}
\includegraphics[width=3.0in]
{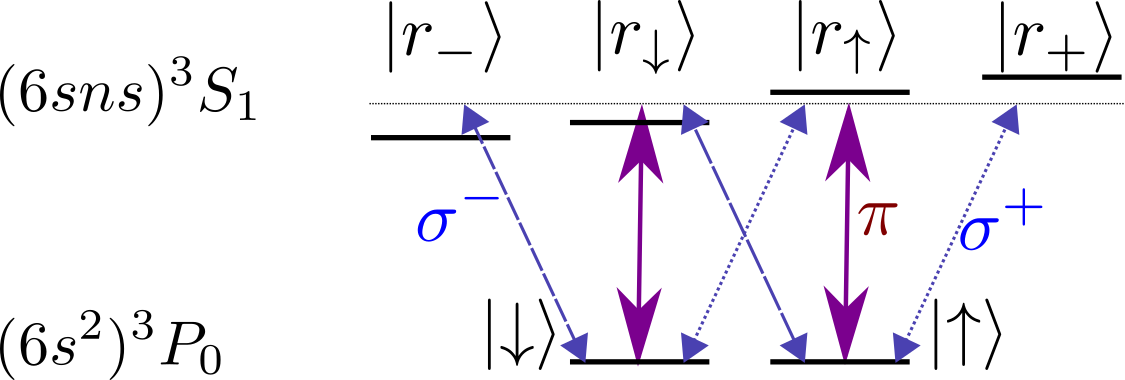}
\caption{Schematic of polarization impurity in the laser field for the gate. Purple arrows indicate the desired transitions, while the blue arrows show transitions induced by unwanted $\sigma^\pm$-polarized laser fields. The laser field has a certain frequency so that the subfields shown are tuned to a virtual level denoted by the dashed line because the Zeeman splitting between $\lvert\uparrow\rangle$ and $\lvert\downarrow\rangle$ is negligible. The detunings are $\Delta_{\uparrow,\downarrow}$ at $\lvert r_{\uparrow,\downarrow}\rangle$, and $\Delta_{\uparrow}-\Delta_{\text{Z}}$ and $\Delta_{\downarrow}+\Delta_{\text{Z}}$ at $\lvert r_\pm\rangle$, respectively. The numerical results shown are based on $\Delta_\downarrow=-\Delta_\uparrow=\Delta_{\text{Z}}/2$ at which $\lvert r_\pm\rangle$ are both more detuned so as to suppress errors.    \label{figure3} }
\end{figure}

\section{Polarization impurity of laser fields}
\begin{figure}
\includegraphics[width=3.30in]
{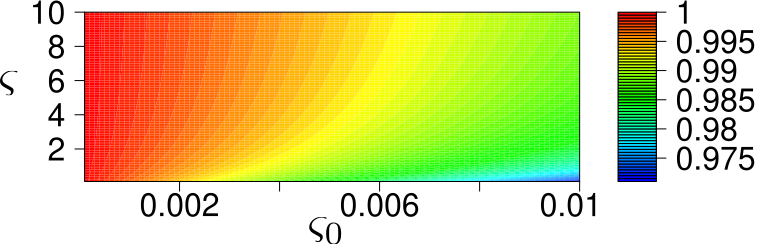}
\caption{Fidelity of the gate with the pulse in Fig.~\ref{figure2}(a) when there are polarization impurity in the laser characterized by $\varsigma_0$ and $\varsigma$, where $\varsigma_0$ is the intensity ratio between the wrong field and the desired field and $\varsigma_0$ is the intensity ratio between $\sigma^+$ and $\sigma^-$ polarized fields. Here, $\mathcal{F}>0.999$ with $\varsigma_0<4\times10^{-4}$, $\mathcal{F}>0.99$ with $\varsigma_0<3\times10^{-3}$, and the maximal and minimal $\mathcal{F}$ are $0.9999$ and $0.9712$ with $(\varsigma_0,\varsigma)$ equal to $(10^{-4},10)$ and $(0.01,0.1)$, respectively.  \label{figure4} }
\end{figure}

The laser field can have impure polarizations leading to unwanted transitions as in Fig.~\ref{figure2} so that the gate map in Eq.~(\ref{Gate01}) becomes a 4$\times$4 matrix where the relevant~\cite{Shi2020} diagonal elements give the gate map diag$\{a e^{i\alpha},b e^{i(\alpha+\beta)/2},b e^{i(\alpha+\beta)/2},-c e^{i(\beta+\epsilon)} \}$ with $a,b,c<1$ and $\epsilon$ is a residual phase. The gate map becomes $\mathscr{U}=$diag$\{a,b,b,-c e^{i\epsilon} \}$ with single-qubit gates, where the difference from the ideal gate $U=$diag$\{1,1,1,-1 \}$ derived from Eq.~(\ref{Gate01}) by single-qubit gates can be characterized by the fidelity~\cite{Pedersen2007} $\mathcal{F} = \left[|\text{Tr}(\hat{U}^\dag \hat{\mathscr{U}})|^2 + \text{Tr}(\hat{U}^\dag\hat{\mathscr{U}}\hat{\mathscr{U}}^\dag\hat{U} )\right]/20$.
If the power ratio between the $\pi$, $\sigma^+$, and $\sigma^-$ polarized fields is $1: \varsigma_0\varsigma/(1+\varsigma):\varsigma_0/(1+\varsigma)$, where $\varsigma_0$ is the intensity ratio between the wrong field and the desired field, then the rates for the undesired transitions denoted by the dashed and dotted arrows in Fig.~\ref{figure3} can be characterized by angular momentum selection rules~\cite{DASteck}, based on which we have simulated the gate fidelity shown in Fig.~\ref{figure4}. Two features appear in Fig.~\ref{figure4}. First, $\mathcal{F}$ is still quite large in the presence of polarization impurity, and $\mathcal{F}>0.999$ when $\varsigma_0<0.0004$. This suggests that the gate can attain a high fidelity in practical implementation for the intensity ratio between the wrong field and the desired field can be made as small as $10^{-4}$ as in the experiment of Ref.~\cite{Dorantes-thesis}. Second, the fidelity shows an unequal dependence on the ratio between the $\sigma^+$ and $\sigma^-$ polarized fields, and it appears that the error is smaller if the wrong polarization is mainly $\sigma^+$. This is because as shown in Fig.~\ref{figure2}(b-d) the final phase in the input state $\lvert \uparrow\uparrow\rangle$ has a more pronounced value for determining the final $\pi$ phase of the gate map. The $\sigma^-$ transition can induce an undesired transition from $\lvert \uparrow\rangle$ to $\lvert r_\downarrow\rangle$, which does more harm to the dynamics of $\lvert \uparrow\rangle$ compared to the $\sigma^+$ transition that induces an undesired transition from $\lvert \uparrow\rangle$ to $\lvert r_+\rangle$ because the detuning $\Delta_\text{Z}/2$ for $\lvert \uparrow\rangle\rightarrow\lvert r_\downarrow\rangle$ is three times smaller than the detuning $-3\Delta_\text{Z}/2$ for $\lvert \uparrow\rangle\rightarrow\lvert r_+\rangle$.

It is in principle feasible to suppress the gate errors from the polarization impurity of the laser fields. The impure polarization is mainly from the misalignment of the quantum axis (usually specified by the magnetic field) and the propagation directions of the laser fields~\cite{Dorantes-thesis}, both of which are fixed during the gate sequence. So the laser polarization impurity can be determined~\cite{PhysRevLett.119.180503,PhysRevLett.119.180504,Dorantes-thesis}. Therefore, with a certain $\varsigma_0$ and $\varsigma$, it is in principle possible to find optimal pulses~\cite{Saffman2020,PhysRevA.105.042404,Jandura2022,PhysRevApplied.18.044042,Jandura2023,PhysRevResearch.5.033052,Sola2023a,Sola2023b,Li2023,Sun2023,PhysRevLett.131.110601,HHJen2023,PRXQuantum.4.040333} for maximizing the gate fidelity. But even without doing so, Fig.~\ref{figure4} indicates that our gate protocol based on easily attainable linearly polarized laser can yield high fidelity because $\varsigma_0$ can be made as small as $10^{-4}$ in the experiment of Ref.~\cite{Dorantes-thesis}. Given the fact that the polarization purity in~\cite{Dorantes-thesis} was achieved more than seven years ago with circularly polarized fields, it is reasonable to assume that technology has been advancing and a higher purity is possible with the gate here since linear polarization is easier to prepare.

\section{ Discussion and conclusions}
We have studied a fast CZ-like gate with nuclear spins in a weak magnetic field by taking $^{171}$Yb as an example for it has the simplest nuclear spin, $I=1/2$, which allows relatively easier manipulation in experiments~\cite{PRXQuantum.4.030337,PhysRevX.13.041034,PhysRevX.13.041035}. Besides, the clock state of $^{171}$Yb possessing only two nuclear-spin Zeeman substates results in that there is no other nuclear spin states for the population to leak to~\cite{footnote3} as shown in Fig.~\ref{figure3}. For atoms with larger $I$, the theory is applicable by strong Stark shifts to shift away nearby nuclear-spin Zeeman substates as in~\cite{Barnes2022} so as to suppress population leakage outside of the qubit-state space, and faster gates are possible.

In summary, a global laser pulse with a smooth modulation of phase can induce a CZ-like gate between two atoms in their nuclear-spin qubit states when $\Delta_{\text{Z}}<\Omega$, where $\Delta_{\text{Z}}$ is the Zeeman-splitting-dominated frequency difference for the clock-Rydberg transitions of the two nuclear-spin qubit states and $\Omega$ is the maximal Rabi frequency. The minimal $\Delta_{\text{Z}}$ is about $0.6\Omega$ for realizing such a gate via linearly polarized laser fields where the gate duration is about $2.2\pi/\Delta_{\text{Z}}$, which sets the speed limit for the gate in an anticipated practically useful quantum computer based on large-scale nuclear-spin memories under a weak B-field. The gate can attain a high fidelity with lasers of experimentally affordable polarization purity~\cite{PhysRevLett.119.180503,PhysRevLett.119.180504,Dorantes-thesis}.

\section*{ACKNOWLEDGMENTS}
The research leading to the results here has received funding
from the National Natural Science Foundation of China under Grants No. 12074300 and No. 11805146, the Innovation Program for Quantum Science and Technology 2021ZD0302100, and the Natural Science Basic Research plan in Shaanxi Province of China under Grant No. 2020JQ-287.

%

\end{document}